\documentclass[amsmath,amssymb,showpacs,aps,prd,superscriptaddress,floatfix,nofootinbib]{revtex4-2}

\usepackage{amsmath, amssymb, epsfig, color}
\usepackage{lineno}
\usepackage{graphicx}
\usepackage{dcolumn}
\usepackage{bm}

\usepackage{aas_macros}
\DeclareMathAlphabet{\mathpzc}{OT1}{pzc}{m}{it}

\begin{document}


\preprint{APS/123-QED}

\title{Manuscript Title:\\On the nature of radio-wave radiation from particle cascades}

\author{Clancy W.\ James}
\email{clancy.james@curtin.edu.au}
\affiliation{
 International Centre for Radio Astronomy Research, Curtin University, Bentley, WA 6102, Australia
}%

\date{\today}

\begin{abstract}
The nature of the radio-wave radiation generated by particle cascades in both the Earth's atmosphere and dense media such as ice has, historically, been much debated. This situation changed in the early 2010's, with the community converging on the common terminology of ``geomagnetic'' and ``Askaryan'' radiation to describe the two emission mechanisms. However, this convergence arose from discussions at various conferences and workshops, and was ultimately reached through agreement between simulation codes and experimental measurements.
In this article therefore, I use relatively simple geometrical arguments, and a minimum of calculations based on single particle tracks, to explain the nature of radiation from extensive air showers (EAS) and cascades in dense media such as ice. I identify well-determined frequency regimes where the radiation from the Askaryan effect will be bremsstrahlung-like and Cherenkov-like, being respectively below/above 1\,GHz in EAS and 100\,MHz in dense media; and where geomagnetic emission will be transverse-current-like and where it will resemble synchrotron radiation, respectively below/above a few GHz in EAS, depending on the height of cascade development. I suggest how these transitions in the nature of the emission may be experimentally observed.
\end{abstract}

\maketitle

\section{Introduction}
\label{sec:intro}

The $21^{\rm st}$ century has seen a renaissance in the use of radio emission to study high-energy particle cascades. Broad uses include the study of extensive air showers in the Earth's atmosphere \cite{LOFAR_detecting,AERA,TUNKAREX,GRAND2020}, searches for neutrinos in deep Arctic and Antarctic ice \cite{ANITA_results_3rd,ARA,ARIANNA}, and searches for particle cascades of both varieties in the lunar regolith \cite{LUNASKA_Parkes,NuMoon_Westerbork}. An excellent summary of recent experimental activity is given in Ref.\ \cite{2017PrPNP..93....1S}.

The radio emission of such cascades is understood as being due to two fundamental mechanisms. The Askaryan effect explains how an excess of negative charge is built up at the shower front, leading to coherent radiation with axially symmetric polarisation \cite{Askaryan62}. Geomagnetic emission occurs due to the deflection of charged particles, particularly $e^\pm$, in the Earth's magnetic field, yielding radiation with polarisation in the direction of the Lorentz force \cite{1966RSPSA.289..206K}. For extensive air showers (EAS) in the Earth's atmosphere, the geomagnetic mechanism is generally dominant \cite{1967Natur.215..267A}, while in the case of dense media (ice and regolith), it will be solely due to the Askaryan effect. For further discussion on how cascade properties influence the nature of the radiation, I recommend Ref.\ \cite{Huege_Review} and Ref.\ \cite{2006PhRvD..74b3007A} for EAS and dense media respectively.

The resurgence in the technique has been due to two factors. Experimentally, the advent of digital radio astronomy provided the necessary fast signal processing for detecting the nanosecond-scale emission from these events, with pathfinding experiments RICE \cite{RICE}, CODALEMA \cite{CODALEMA}, LOPES \cite{LOPES}, and Parkes \cite{Parkes_1995}. 
The second factor is an increased theoretical understanding of the emission mechanism. While many experiments have now measured radio-emission from EAS,\footnote{To date, there has been no firm identification of radio emission from a cosmic particle cascade in a dense medium.}, there were initially competing explanations for these observations. Geomagnetic emission was modeled `microscopically' as the sum of synchrotron radiation from individual particles, or `macroscopically' as radiation from the current resulting from the motion of those particles. While these two approaches should yield the same result, it was not until 2010 that these two models could be reconciled \cite{2012NIMPA.662S.179H}. The breakthrough came with the realization of the importance of finite track lengths and the associated bremsstrahlung-like emission in the Askaryan effect \cite{endpoints}, which had previously been understood as coherent Cherenkov radiation.

The current status of the field is excellent. The ZHS code \citep{ZHS_1992}, and its extensions to model hadronic cascades \citep{ZHS_hadronic}, has long been able to model the radio emission from cascades in dense media such as ice, which can be reproduced in laboratory measurements using particle accelerators \citep{SLAC_sand,SLAC_ice}. Two numerical codes -- CoREAS \cite{CoREAS} and ZHAireS \cite{ZHAireS} --- produce mutually consistent results that agree with experimental observations of EAS \cite{LOFAR_Xmax,LOFAR_polarisation,AugerRadioEnergyPRL,ANITA_UHECR,ARIANNA_CR}, and accelerator measurements with magnetic fields \cite{SLAC_magnetic,SLACT510_detailed}. Semi-analytic methods such as EVA \cite{EVA} and MGMR3D \cite{MGMR3D} produce emission profiles in a much shorter time interval with little loss of accuracy. The community has also converged on the common terminology of `geomagnetic' and `Askaryan' radiation to describe the two emission mechanisms, while the terminology of `transverse current', `geosynchrotron', `bremsstrahlung', and `Cherenkov' has mostly been dropped as being too reductive.

This begs the question -- why revisit a solved problem?

There are two good reasons to do so. Firstly, our current understanding of the radio-emission from particle cascades was arrived at only through lively discussions at conferences --- particularly the Acoustic and Radio EeV Neutrino detection Activities (ARENA) workshop series --- and many explanations have so-far remained unpublished. Furthermore, while codes such as CoREAS and ZHAireS `just work', they are computationally expensive black boxes, and may need to be tuned to new physical situations. Fast semi-analytic calculations such as MGMR3D \cite{MGMR3D} are therefore very useful, but their accuracy relies upon an understanding of the properties of an EAS leading to radio emission. The first goal of this article therefore is to elucidate the qualitative nature of the radio-wave radiation from particle cascades. For simplicity, I consider only individual particle tracks, rather than the more-complicated case of entire particle cascades.

Secondly, consensus was ultimately reached through the agreement between simulation codes and experimental measurements, most of which have been conducted for extensive air showers in the 30--100\,MHz range. Yet radiation phenomenology is frequency-dependent. Furthermore, while predictions for accelerator experiments mimicking particle cascades in dense media show reasonable agreement with measurements \cite{SLAC_sand,SLAC_ice,SLAC_magnetic,SLACT510_detailed}, no radio emission from a cosmic particle interacting in a medium other than the Earth's atmosphere has ever been identified. The major goal of this article therefore is to demonstrate how the nature of the radiation will change as a function of frequency and the interaction medium, and why. This is analyzed for the case of the Askaryan effect using straight particle tracks in  Sec.~\ref{sec:Askaryan}, while the slightly more complicated case of geomagnetic emission and curved particle tracks is discussed in Sec.~\ref{sec:synchrotron}.

\section{Media}

Current experimental activities in the radio-detection of high-energy particles predominantly use the Earth's atmosphere and Arctic/Antarctic ice \cite{2017PrPNP..93....1S}. A smaller number of experiments have also used the lunar regolith as a target medium, with salt domes and permafrost also being proposed for future experiments. These latter media are sufficiently similar to ice in terms of density and refractive index that results obtained in ice can be adapted to them with small scaling factors \cite{2006PhRvD..74b3007A}. Rather than simulate a large number of potential media, here I stick simply to two cases: the Earth's atmosphere, and ice. 

It is worth noting however that new and unique target media could feasibly be proposed, e.g.\ Jupiter's atmosphere, where simple scaling of radiation properties from those observed in the Earth's atmosphere fails \cite{2016ApJ...825..129B}. It is the goal of this work to impart a sufficient understanding of the nature of radiation from particle cascades that sensible estimates could be made for new media without the need for detailed and unique simulations.

\begin{table}[b]
\caption{\label{tab:media} Characteristic parameters for cascades in the atmosphere and in ice \cite{2012PhRvD..86a0001B}.
}.
\begin{ruledtabular}
\begin{tabular}{l | c c c c c }
& $E_{\rm crit}$ & $\chi_0$  & $ \rho $ & $\chi_0/\rho$ & $n$ \\
& MeV & g cm$^{-2}$ & g cm$^{-3}$ & m &  \\
\hline
Atmosphere (sea level) & 87.92 & 36.62 & 0.0012 & 305 & 1.0003  \\
Ice & 78.6 & 36.08 & 0.918 & 0.4 & 1.8 \\
\end{tabular}
\end{ruledtabular}
\end{table}

Relevant properties of ice and sea-level air are given in Tab.~\ref{tab:media}, taken from Ref.\ \cite{2012PhRvD..86a0001B}. Atmospheric properties are scaled with height using the US standard atmosphere \cite{1976ussa.rept}, as implemented in \textsc{Python} by Ref.\ \cite{PDAS}. By default, I use values at sea level, unless noted otherwise.

Quantities of particular note are the refractive index $n$, and density $\rho$. The latter determines the characteristic length of particle tracks in a cascade through the radiation length $\chi_0$. Formally, this is the distance over which an $e^{\pm}$ will retain on average $1/e$ of its energy. In the atmosphere, where density $\rho$ decreases with altitude, the critical energy $E_{\rm crit}$ and $\chi_0$ remain constant, such that $\chi_0/\rho$ increases with altitude, while $n-1$ decreases proportionally to $\rho$.

For quantitative calculations, simulation programs approximate particle trajectories as a sequence of straight lines, called `tracks'. For accurate results, these tracks must be at least as small as a radiation length \cite{AMZCalculationMethods}. Here, the radiation length will be used as the characteristic distance over which a particle trajectory can be considered to be straight for phenomenological arguments.

\section{Askaryan radiation}
\label{sec:Askaryan}

The Askaryan effect is due to the charge excess built up primarily through knock-on electrons in the shower front \cite{Askaryan62,Askaryan65}. The radiation is regularly identified as coherent Cherenkov radiation, e.g.\ Ref.\ \cite{2018PhRvD..98c0001T}, since its origin is in a charge moving superluminally (velocity $v=\beta c$, i.e.\ $\beta \equiv v/c$) through a medium with refractive index $n$, and it exhibits a peak at the Cherenkov angle, $\theta_C = \arccos (\beta n)^{-1}$.

What is commonly forgotten however is that the Askaryan effect is fundamentally significant because it gives a mechanism by which a particle cascade can emit coherently, where otherwise emission from positive and negative charges would cancel. The nature of that emission, however, depends on the nature of the behavior of those particles. In their 1962 paper, Askaryan suggested that this mechanism would allow coherent bremsstrahlung, Cherenkov radiation, and transition radiation. Indeed, the phenomena of superluminal motion leading to the coherent addition of radiation at the Cherenkov angle --- sometimes known as ``Cherenkov effects'' --- is more general than the specific emission of Cherenkov radiation itself (Refs.~\cite{ZHAireS,Scholten_cherenkov_description} and Ref.~\cite{2006PhRvD..74b3007A} give good discussions of coherency in EAS and dense media respectively). However, by 1965 only Cherenkov radiation remained in the terminology.

The radiation observed by Cherenkov \cite{Cherenkov37} was explained by I.~Frank and I.~Tamm (Ref.\ \cite{FrankTamm37}) in a calculation considering a particle moving for an infinite distance through a uniform medium. Tamm (Ref.\ \cite{Tamm39}) later analyzed the case of a particle moving for a finite distance (``the Tamm problem''), and identified two components to the radiation --- one identified as Cherenkov radiation due to the motion, and a correction due to bremsstrahlung at the ends of the track.

There is a long history of debate regarding the relative influences of bremsstrahlung and Cherenkov radiation in the Tamm problem, as discussed by e.g.\ Ref.~\cite{Afanasiev_1999} and Ref.~\cite{BuniyRalston}.
In particular, radiation from the Askaryan effect in particle cascades being predominantly Cherenkov in nature was challenged by Ref.~\cite{endpoints}, who provided a derivation of radiation that explicitly depended on the implied particle acceleration at the beginning and end of particle tracks, i.e.\ ``endpoints''. Macroscopically, this component of the radiation arises due to the rise and fall of the negative charge excess in a cascade, and is analogous to the bremsstrahlung originally suggested by Askaryan. As such, it was orthogonal to the calculation of Frank and Tamm for Cherenkov radiation, which relied only on particle \emph{motion}.

The implementation of the endpoints calculation into the air-shower code \textsc{CORSIKA} \cite{CORSIKA_1998} as \textsc{CoREAS} \cite{CoREAS} enabled the successful reconstruction of cosmic ray events by LOFAR \cite{LOFAR_Xmax}. Furthermore, it was realized that the formula for radio Cherenkov radiation published in Ref.\ \cite{ZHS_1992} in fact applied to all radiation processes \citep{Exact}, leading to its implementation in \textsc{ AIRES}\footnote{http://aires.fisica.unlp.edu.ar/} to produce the \textsc{ZHAireS} code \cite{ZHAireS}. As a result, the EAS community has largely dropped the ``Cherenkov'' nomenclature and refers simply to the component of radiation due to the charge-excess mechanism as ``Askaryan radiation''.

Yet, at optical--UV wavelengths, the emission detected by imaging atmospheric Cherenkov telescopes (IACTs) is clearly Cherenkov radiation. How and why does the nature of the radiation change with frequency?

\begin{figure}
    \centering
    \includegraphics[width=0.5\textwidth]{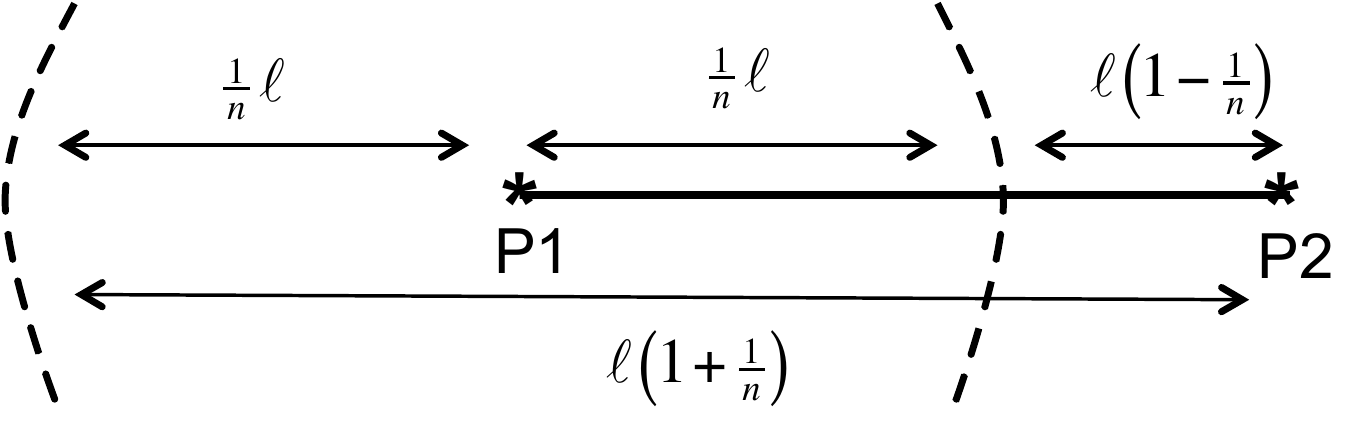}
    \caption{Sketch of the electromagnetic shock front (dotted line) from a particle initially accelerated at P1 in a medium with refractive index $n$ and having travelled a distance $\ell$ to P2 with uniform velocity $\beta=1$ (solid line).}
    \label{fig:formation}
\end{figure}

The answer is given in Fig.~\ref{fig:formation}. Consider a particle which is initially created/accelerated at P1, and travels a distance $\ell$ to P2. Assuming $\beta=1$, the time taken for this motion will have been $\Delta t=\ell/c$. The shock of the particle's sudden acceleration at P1 will have traveled a distance $\Delta t c/n = \ell/n$ from P1, shown as the dotted lines in Fig.~\ref{fig:formation}. The distance from P2 to the right-most leading edge of the shock front from P1 will be $\ell (1-1/n)$. If this distance is less than the wavelength in the medium, $\lambda_n \equiv \lambda/n$, any radiation at P2 will not be separable from the P1 shock, and the emission will be at least partially bremsstrahlung-like. If this distance is greater than $\lambda_n$, then events at P2 will appear to be separated from P1. That is, the particle's motion will have decoupled it from the initial acceleration, since information about that acceleration will have lagged behind the particle.\footnote{I use the term `decoupled' here to mean that the total power emitted will be a linear combination of the separate processes --- the fields themselves will, in general, show an interference pattern reflecting the entirety of the particle motion.} This would allow, for instance, canonical Cherenkov emission to be produced, without consideration of the initial acceleration.

Radiation emitted in the backwards direction (i.e.\ to the left in Fig.~\ref{fig:formation}) will decouple more rapidly however, when  $\lambda_n < \ell(1 + 1/n)$, since the shock front and particle are moving in opposite directions. Thus only deceleration within a very short distance $\ell < \lambda_n (1+1/n)^{-1}$ after an initial acceleration will result in coupled radiation in the backwards direction.

\begin{figure}
    \centering
    \includegraphics[width=0.24\textwidth,trim={3cm 0cm 3cm 0cm},clip=True]{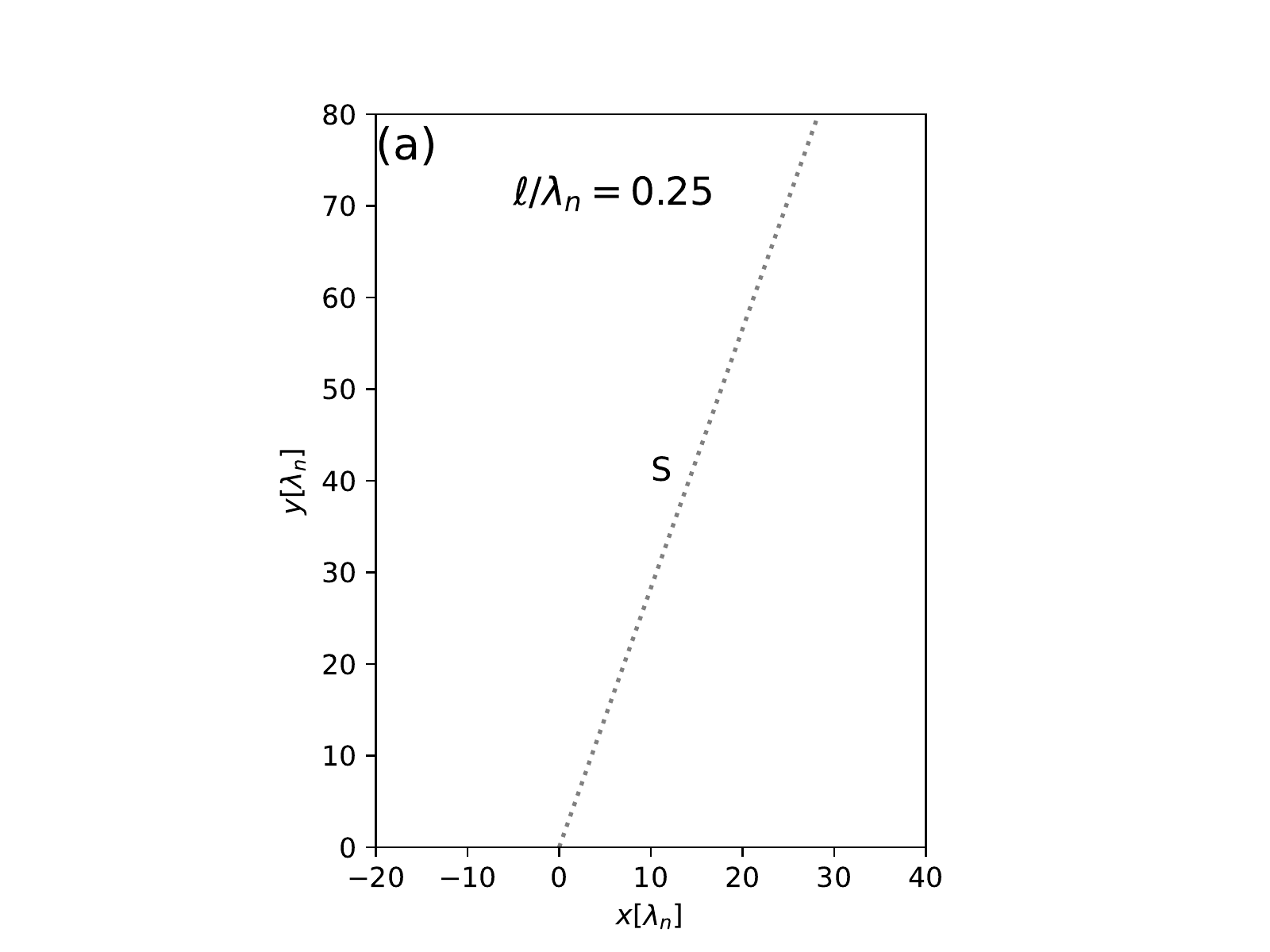}
    \includegraphics[width=0.24\textwidth,trim={3cm 0cm 3cm 0cm},clip=True]{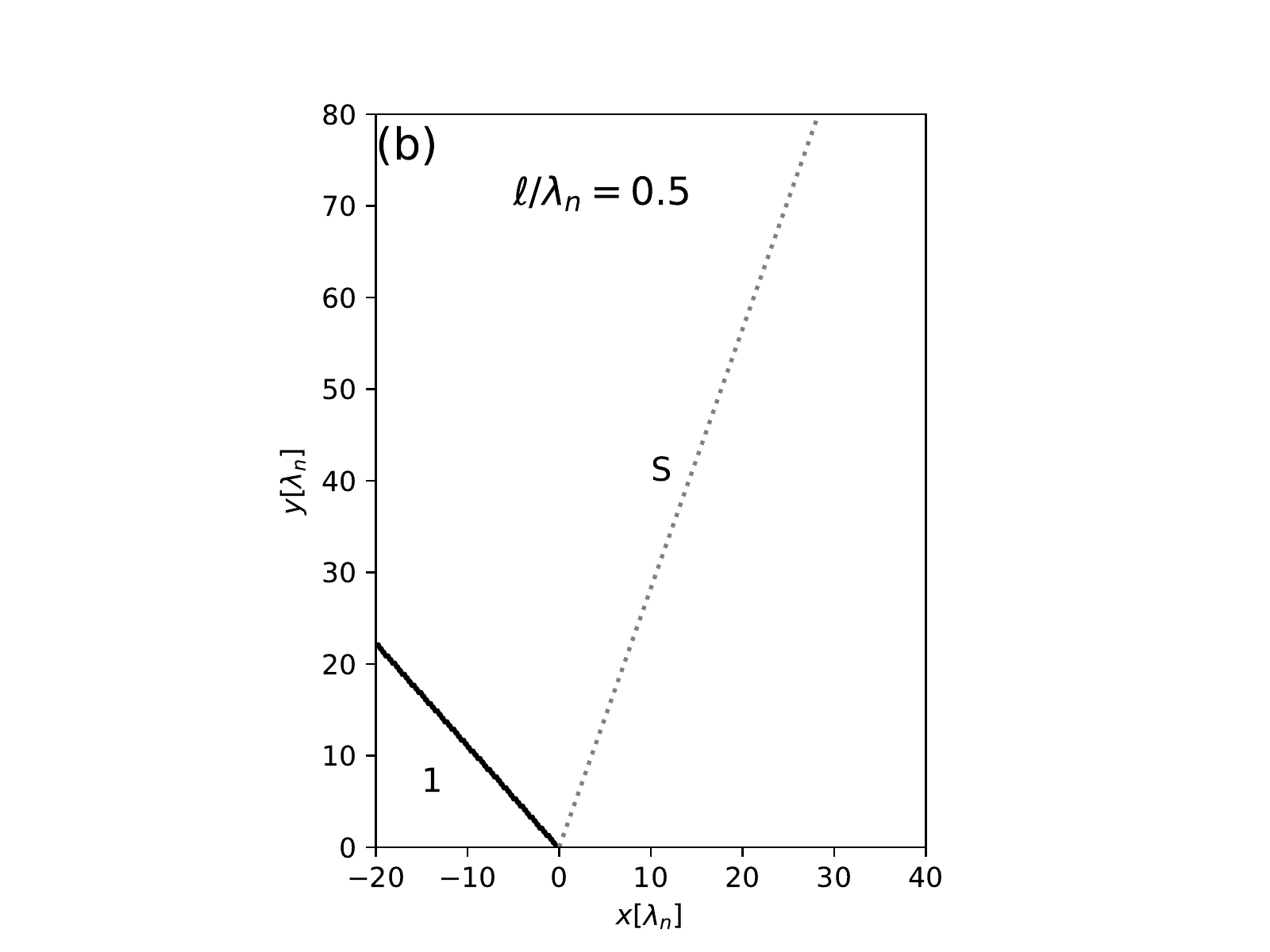}
    \includegraphics[width=0.24\textwidth,trim={3cm 0cm 3cm 0cm},clip=True]{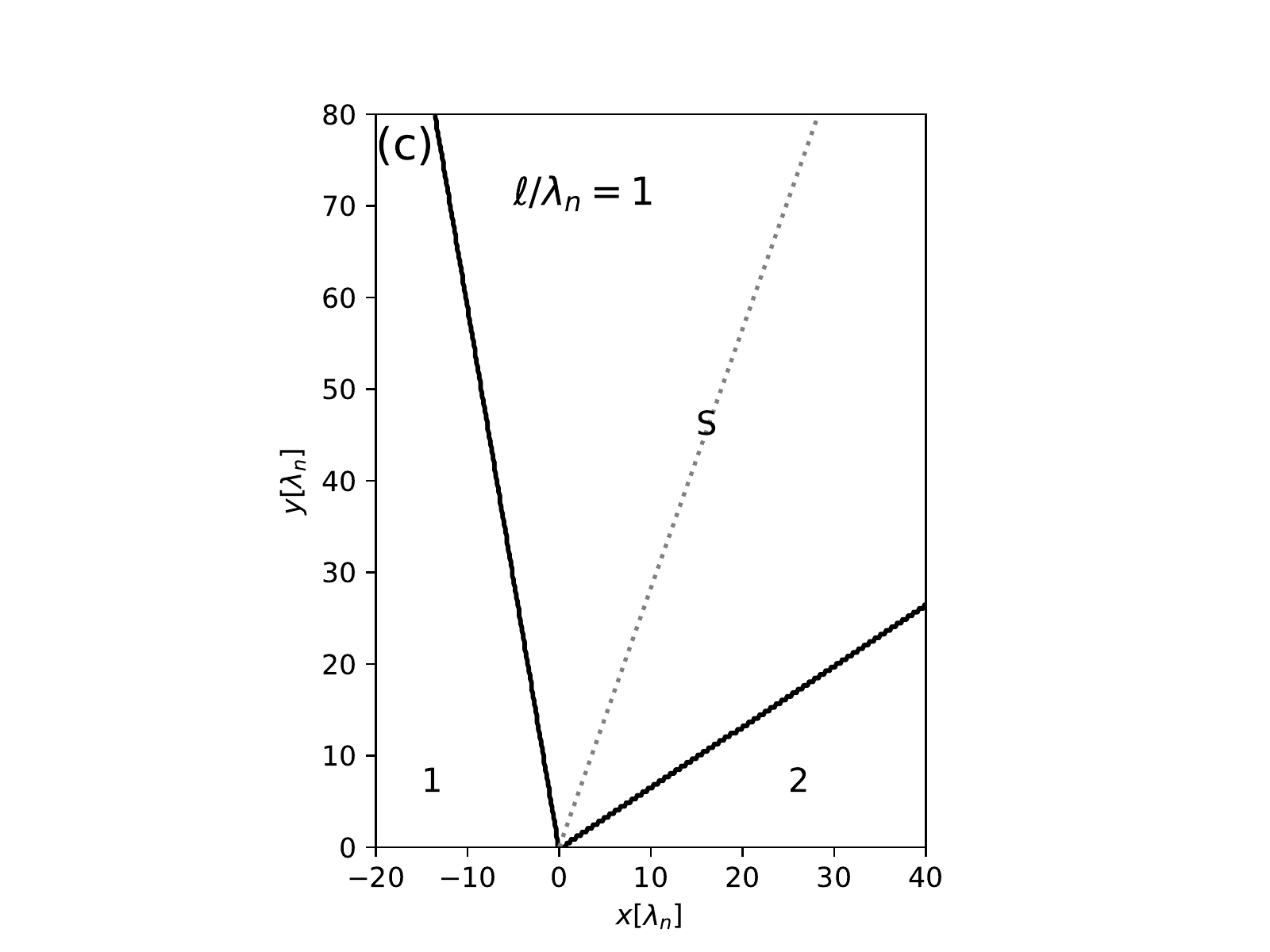}
    \includegraphics[width=0.24\textwidth,trim={3cm 0cm 3cm 0cm},clip=True]{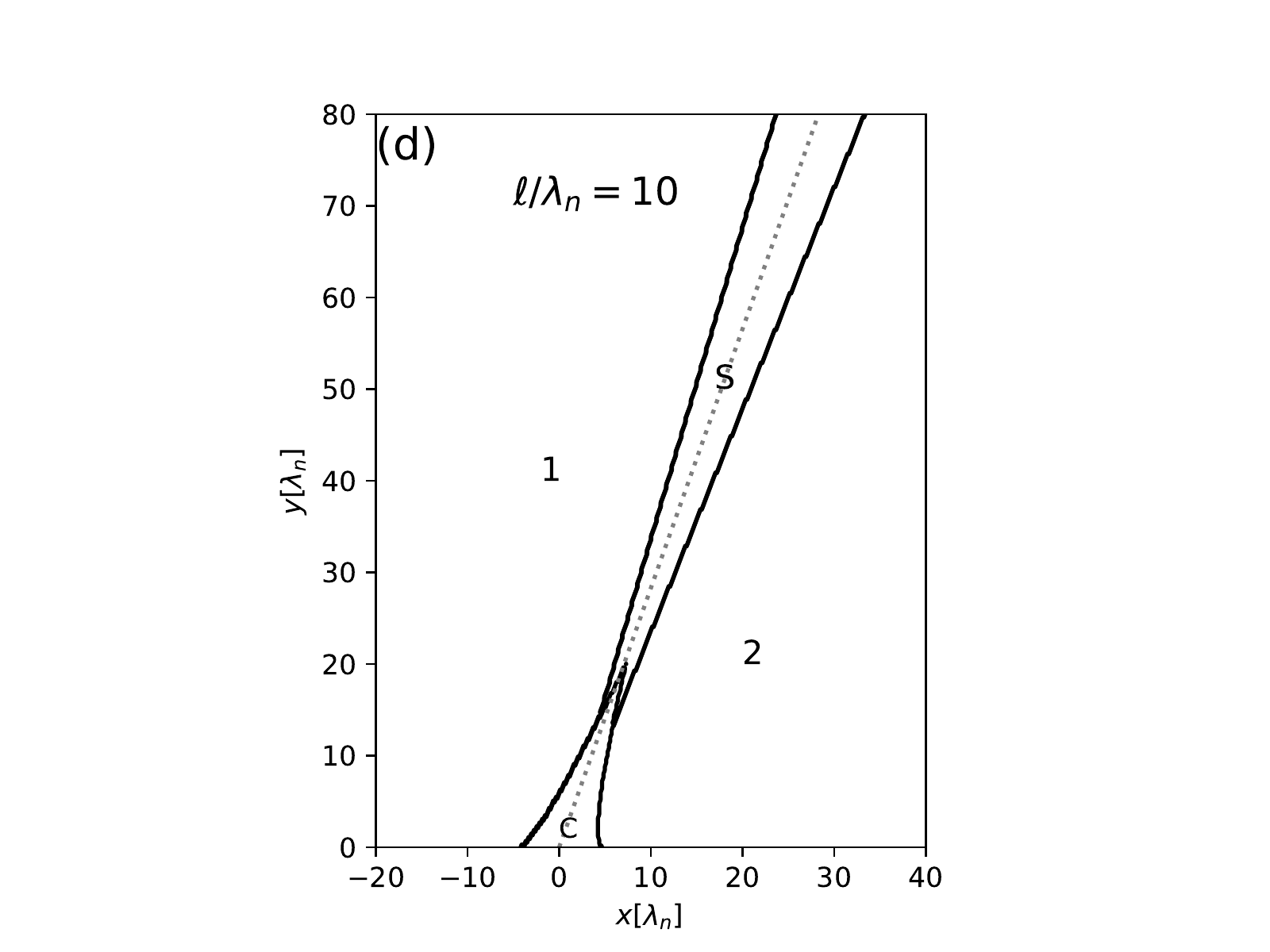}
    \caption{Illustration showing regions in the x--y plane where an observer can distinguish different shocks, i.e.\ arrival times are separated by at least half a period, $T/2=(2 \nu)^{-1}=\lambda/c$, from a particle travelling between P1 and P2 with $\vec{v}=c \hat{x}$ at $y=0$ observed at frequency $\nu$. Each panel shows increasing tracklength $\ell$ in units of wavelength in the medium $\lambda_n=\lambda/n$, centred at $x=0$ (i.e.\ start and stop points P1 and P2 are at $\pm \ell/(2 \lambda_n)$). I use $n=3$ for illustrative clarity, with the Cherenkov angle $\theta_C = \cos^{-1} n^{-1}$ labelled as a grey dashed line. In regions labelled `1' and `2' the shocks from P1 and P2 arrive first and can be distinguished from those at P2 and P1 respectively; in the region labelled `C' the Cherenkov shock front can be distinguished from other shocks; and in region `S' all shocks are simultaneous, i.e.\ no shocks can be distinguished from each other.}
    \label{fig:observer}
\end{figure}

How this radiation is viewed by an observer can be more complicated however, being dependent on the observer location, as well as the nature of the radiation source. In the far-field (Fraunhofer limit, i.e.\ $R \gg \ell^2/\lambda_n$) region, radiation can be described by the observer angle $\theta$, and field intensity must fall with distance $R$ as $R^{-1}$ (and hence total radiated energy falls as $R^{-2}$). However, at nearer distances, there will in-general be zones of finite extent with unique phenomenology. This is illustrated in Fig.~\ref{fig:observer}.

Fig.~\ref{fig:observer}(a) illustrates the situation for short tracks, where $\ell/\lambda_n < n/ (n+1)$. Observers at all $R,\theta$ will view all emission `simultaneously', i.e.\ within a time $\Delta t \lesssim (2 \nu)^{-1}$ for frequency $\nu$. Thus an observer viewing the particle at frequency $\nu$ will detect a single event.

As the track length $\ell/\lambda_n$ increases (from left to right in Fig.~\ref{fig:observer}), emission from P1 and P2 become distinguishable at some angles. Initially, this occurs in the backwards direction only (P1 before P2; Fig.~\ref{fig:observer}(b)), and then also in the forwards direction (P2 before P1; Fig.~\ref{fig:observer}(c)). Near the Cherenkov angle ($\cos \theta_c = (\beta n)^{-1}$) in the far-field, emission from all points in the particle track will always arrive sufficiently simultaneously so as to be indistinguishable, even for long tracks, as shown in Fig.~\ref{fig:observer}(d).

In the near-field (Fresnel region) of long tracks ($\ell/\lambda_n > n/(n-1)$), shown in Fig.~\ref{fig:observer}(d), a region will emerge where emission from the particle track itself, rather than points P1 and P2, becomes distinguishable. This is a classical Cherenkov shock, and shocks from the accelerations at P1 and P2 will arrive later. For an extended discussion of how shocks are viewed by an observer, I suggest Refs.~\cite{Afanasiev_1999} and \cite{BuniyRalston}, with Ref.~\cite{Exact} quantitatively analyzing the effects of different near- and far-field regimes.

An excellent, observer-independent test of the predicted dependence on tracklength can be given by calculating the total emitted radiation energy per frequency, $dW/d\omega$, for given frequencies as a function of tracklength in a given medium. The calculation is performed using three methods:
\begin{itemize}
    
    \item{Total:} the total emission from a particle track is calculated using the ZHS formula of Ref.\ \cite{ZHS_1992}, which has been shown by Ref.\ \cite{Exact} to reproduce the complete radiation from a particle track when sufficiently many track subdivisions are used, and is implemented in the \textsc{ZHAireS} code \cite{ZHAireS}. Up to $10^5$ subdivisions were used for calculations in this work.
    
    \item{Frank-Tamm:} this component is calculated using the Frank-Tamm formula for Cherenkov radiation \cite{FrankTamm37}, which considers only the motion of the particle.
    
    \item{$d \beta/dt$:} The contribution of acceleration (i.e.\ bremsstrahlung) is calculated according to the endpoints formalism of Ref.\ \cite{endpoints}, which considers only the acceleration term, $d \beta/dt$. Note that the \textsc{CoREAS} code, which by default uses the endpoints formalism, reverts to the ZHS formula at angles very close to the Cherenkov angle, in order to capture the total radiated power and avoid discontinuities \cite{CoREAS}.
    
\end{itemize}

\begin{figure}
    \centering
    \includegraphics[width=0.7\textwidth]{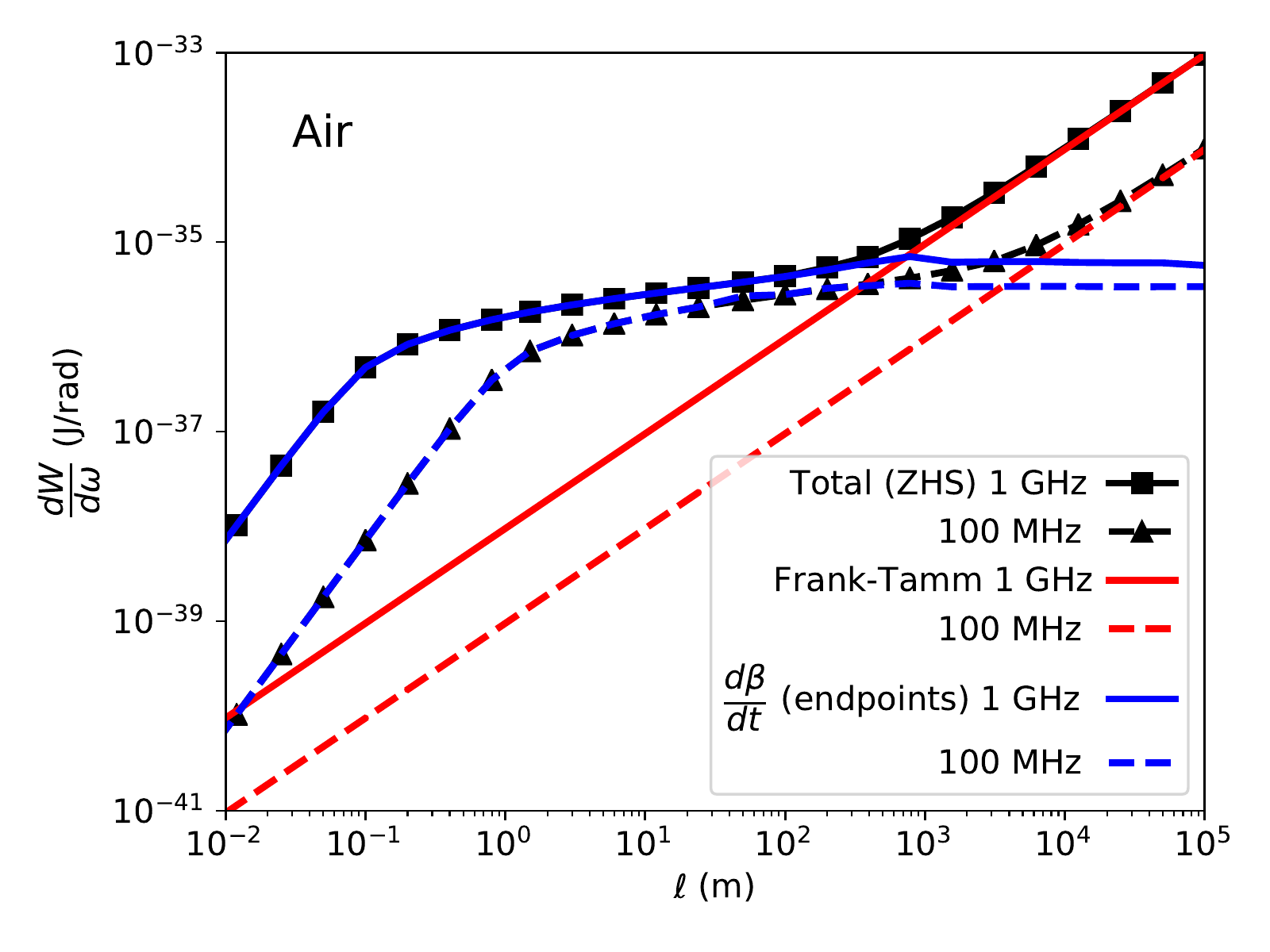}
    \includegraphics[width=0.7\textwidth]{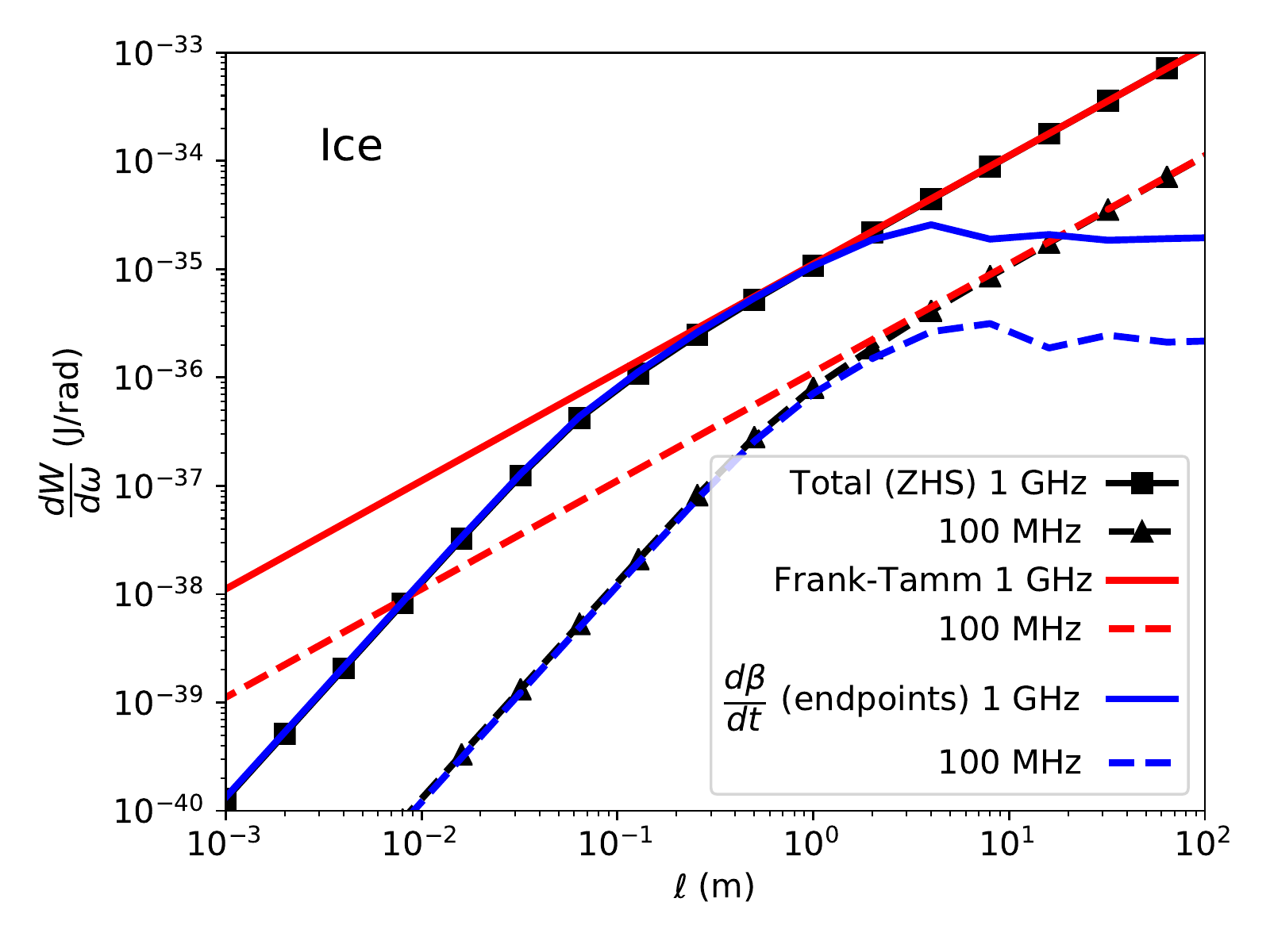}
    \caption{Total radiated energy per angular frequency, $dW/d\omega$, emitted by a particle as a function of $\ell$, i.e.\ undergoing the motion of Fig.~\ref{fig:formation}, at frequencies of 1\,GHz and 100\,MHz, in sea-level air (top) and ice (bottom). Calculations are performed with three different methods, nominally representing the true radiated energy (`total'; ZHS \cite{ZHS_1992}), Cherenkov radiation from particle motion (`Frank-Tamm'; \cite{FrankTamm37}), and bremsstrahlung contributions from particle acceleration (`$d\beta/dt$'; endpoints \cite{endpoints}).}.
    \label{fig:air_ice}
\end{figure}

The Frank-Tamm formula produces $dW/d\omega$ when multiplied by the path length $\ell$, whereas for the `total' and `$d \beta/dt$' contributions, calculations are performed at a large number of angles in the far-field, then the power is integrated over all solid angles. The results in the case of sea-level air and ice at 100\,MHz and 1\,GHz are given in Fig.~\ref{fig:air_ice}.

From the calculations performed in air, three clear regimes emerge. Using estimates at $100$\,MHz, the total power grows with $\ell^2$ until $\ell \approx 1$\,m, as the radiation source is unresolved. This is dipole-like behavior, where the field strength is proportional to the magnitude of the motion. In the range $1 \lesssim \ell \lesssim 10^4$\,m, power remains approximately constant, consistent with bremsstrahlung emission from the start and end points. It is no surprise that both regimes are well-reproduced by the endpoints calculation based on $d\beta/dt$, while the Frank-Tamm prediction under-estimates the total radiated power. For $\ell \gtrsim 10^4$\,m, total power increases proportionally with $\ell$, and is correctly described by the Frank-Tamm formula, i.e.\ it is Cherenkov-like. Thus the $d\beta/dt$ calculation underestimates the total radiated power. Similar behaviour is exhibited at 1\,GHz, with the distance regimes scaled down by a factor of 10 in $\ell$.

It is interesting to note that the breaks at $1$ and $10^4$\,m are well-predicted by setting $\lambda=\ell (1+1/n)$ and $\lambda=\ell(1-1/n)$, respectively predicting breaks at $\ell\approx2$\,m and $10^4$\,m for 100\,MHz emission, and $\ell\approx0.2$\,m and $10^3$\,m for 1\,GHz emission. Given the sea-level radiation length of $\chi_0=305$\,m, this explains why the radio-emission so-far observed in EAS is consistent with the bremsstrahlung-like description, and the Cherenkov nomenclature has been dropped in the EAS community.

The qualitative behavior in ice is markedly different however. The intermediate regime where total power remains constant with $\ell$ is very small at 1\,GHz ($\approx$0.3--1\,m), and negligible at 100\,MHz. This again can be explained through Fig.~\ref{fig:formation}: in ice, the $1/n$ term is small, so that once a track $\ell$ is long enough that backwards-directed radiation from the initial acceleration begins to decouple from radiation from the final deceleration, it need only be a little longer to have forwards-directed radiation decouple as well. Again, the $d \beta/dt$ calculation agrees with the total emission at low $\ell$, while the Frank-Tamm formula agrees with the total emission predicted by ZHS at high $\ell$. Given the radiation length of $\chi_0=39.3$\,cm in ice, this explains why members of the radio-in-ice community tend to continue to use the term ``Cherenkov radiation'' when applied to the Askaryan effect: because it is more consistent with their experiments.

\begin{figure}
    \centering
    \includegraphics[width=0.7\textwidth]{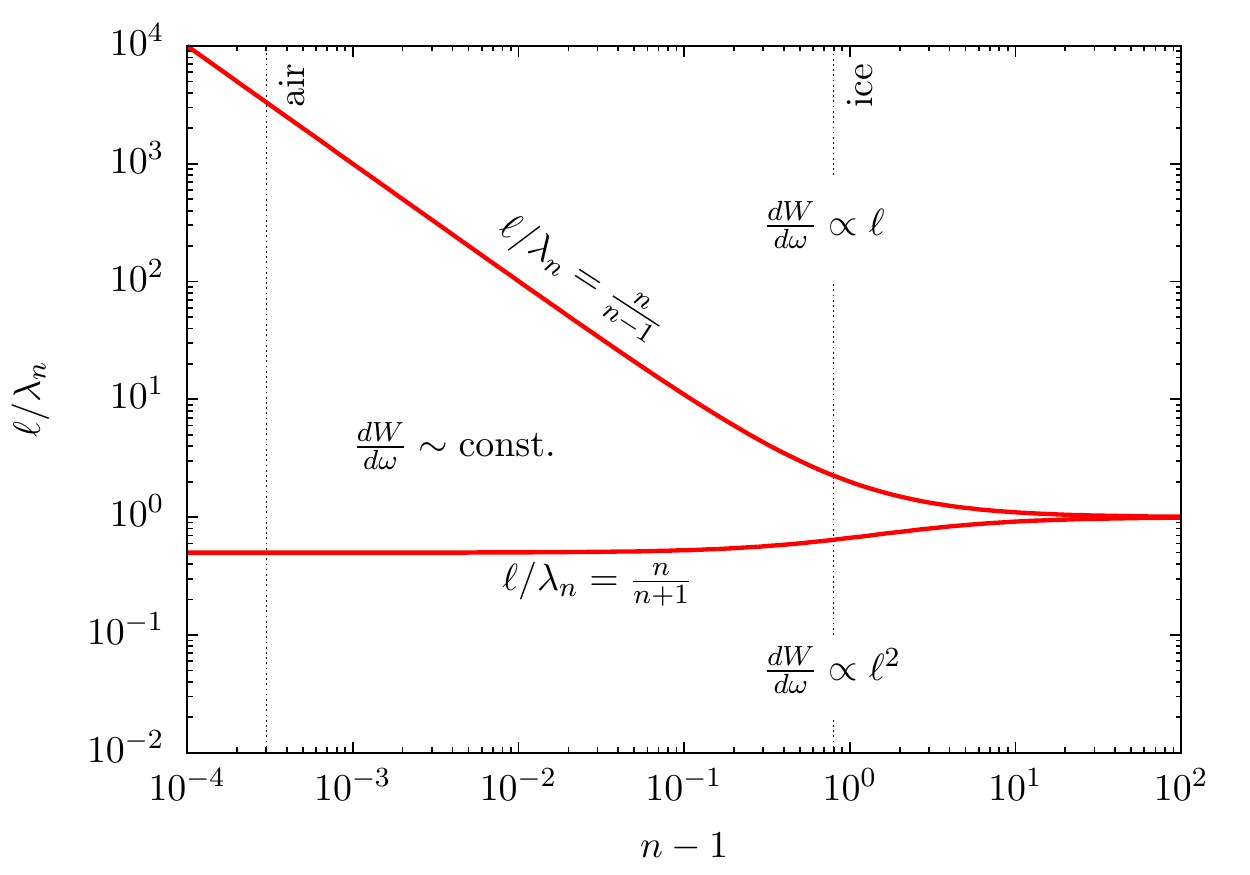}
    \caption{Illustration of the regimes in track-length $\ell$ and medium refractive index $n$ over which the emitted radiation will appear Cherenkov-like (radiated power $dW/d\omega \propto \ell$), bremsstrahlung-like ($dW/d\omega$ constant), or as an unresolved point-source ($dW/d\omega \propto \ell^2$). The values of $n-1$ for air and ice applicable to radio wavelengths are also indicated as vertical lines.}
    \label{fig:regimes}
\end{figure}

The three regimes previously identified can be best expressed when $\ell$ is written in units of the wavelength in the medium, $\lambda_n$. This allows clear demarcations in $\ell/\lambda_n$--$n$ space, which are illustrated in Fig.~\ref{fig:regimes}. For imaging atmospheric Cherenkov telescopes, which observe in the optical and near-UV, $\lambda_n$ is tiny; $\ell/\lambda_n$ is thus very large, and the emission will be firmly within the Cherenkov ($dW/d\omega \propto \ell$) regime described by the Frank-Tamm formula.

Of particular note is that in the atmosphere, $n-1$ scales proportionally with $\rho$. Since radiation length $\chi_0$ does also, the nature of radiation from the Askaryan effect in EAS will be a function only of wavelength, not altitude. This will not however be the case for geomagnetic emission, which is analyzed below.

\begin{figure}
    \centering
    \includegraphics[width=0.7\textwidth]{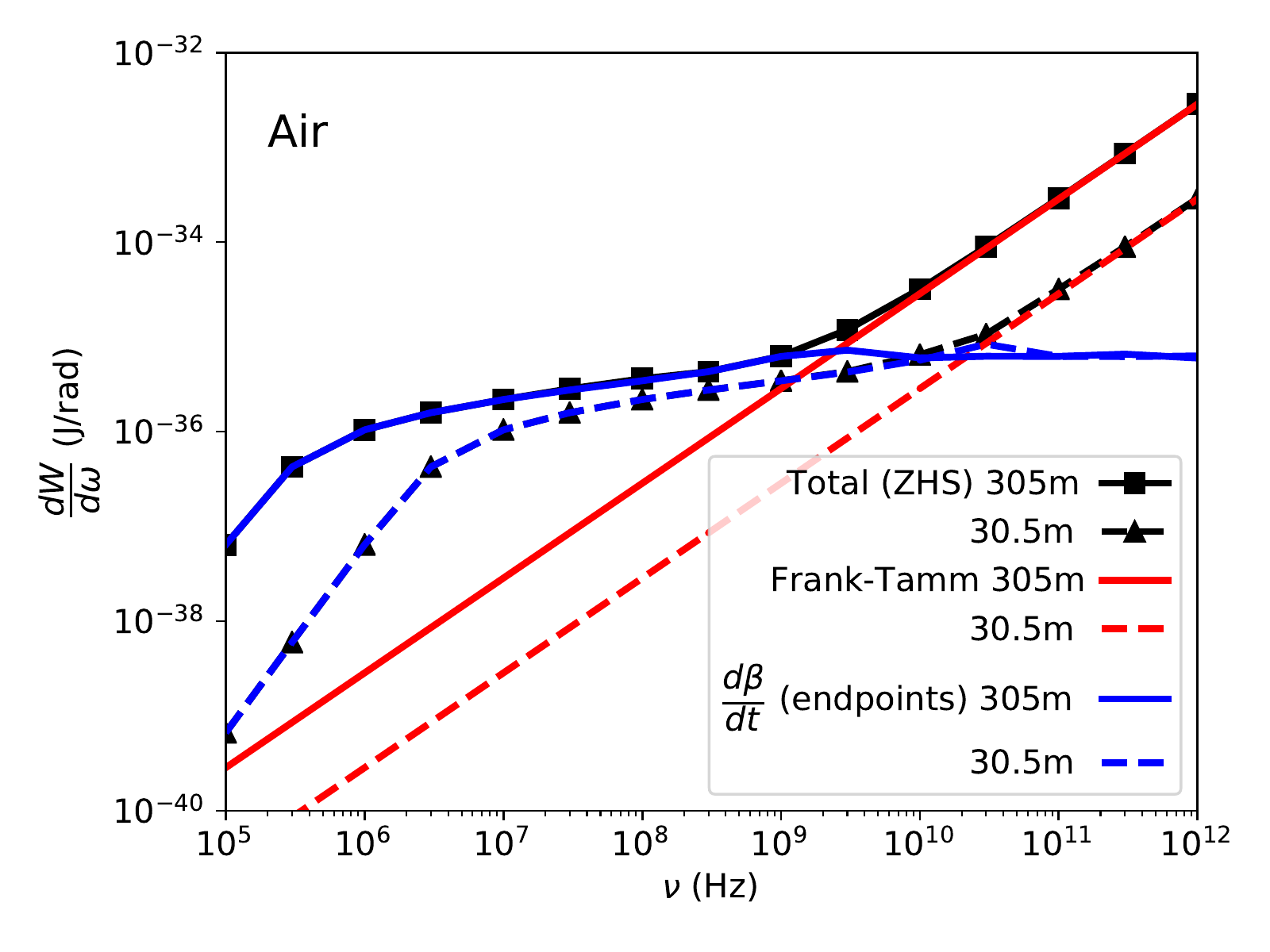}
    \includegraphics[width=0.7\textwidth]{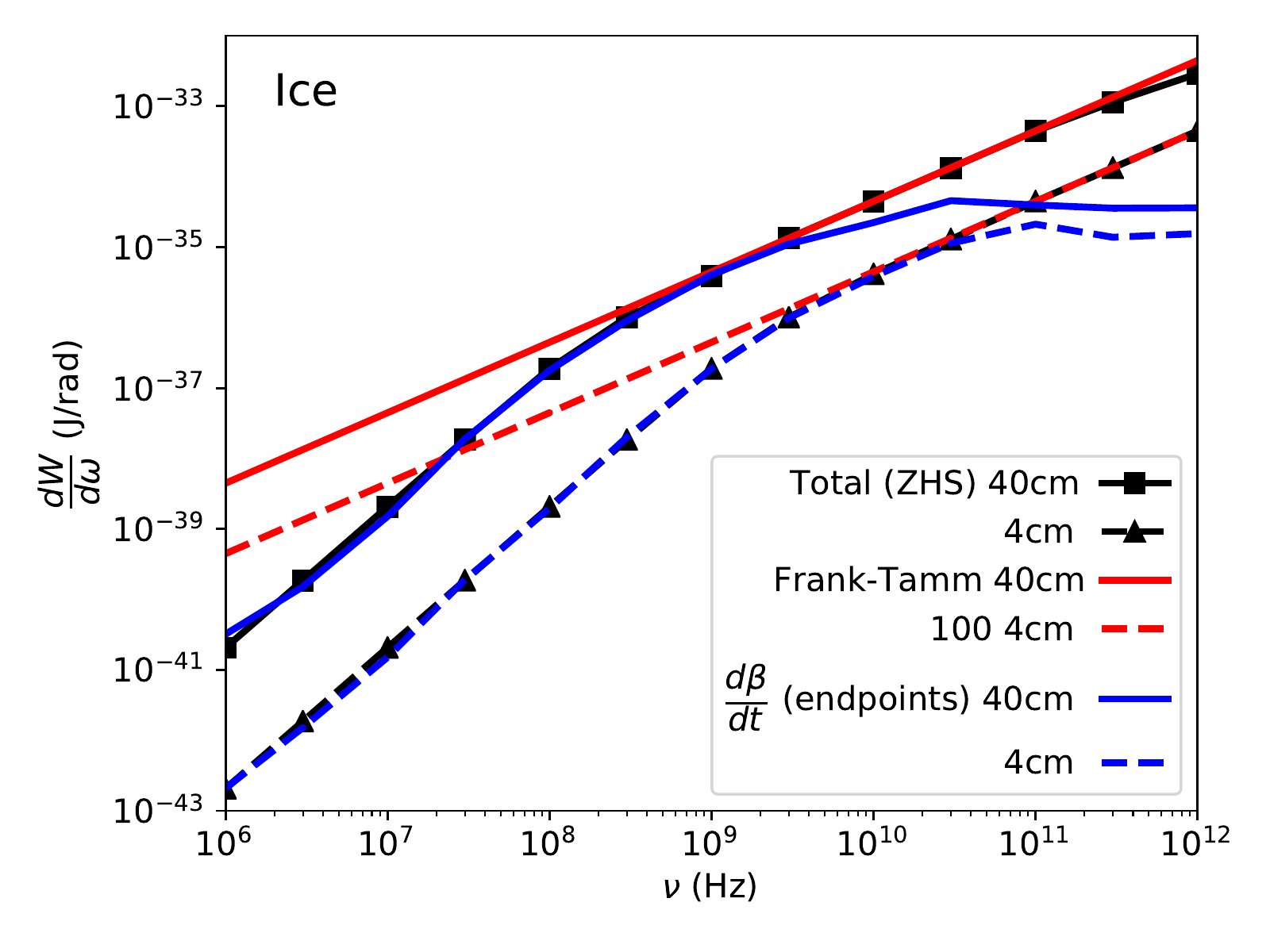}
    \caption{Total radiated energy per angular frequency, $dW/d\omega$, emitted by a particle as a function of frequency $\nu$, i.e.\ undergoing the motion of Fig.~\ref{fig:formation}, at path lengths $\ell$ corresponding to the radiation length $\chi_0/\rho$, and $0.1 \chi_0/\rho$, in sea-level air (top) and ice (bottom). Calculations are performed with three different methods, nominally representing the true radiated energy (`total'; ZHS \cite{ZHS_1992}), Cherenkov radiation from particle motion (`Frank-Tamm'; \cite{FrankTamm37}), and bremsstrahlung contributions from particle acceleration (`$d\beta/dt$'; endpoints \cite{endpoints}).}.
    \label{fig:freq_air_ice}
\end{figure}

The importance of the ratio between track length $\ell$ and wavelength $\lambda_n$ can be seen in Figure~\ref{fig:freq_air_ice}. This is the equivalent of Figure~\ref{fig:air_ice}, albeit with power changing as a function of frequency rather than $\ell$. The behaviour is qualitatively identical, illustrating that it is indeed the ratio $\ell/\lambda_n$ that governs radiative behaviour.

\section{Geomagnetic emission}
\label{sec:synchrotron}

The dominant component of radio emission from extensive air showers is geomagnetic, i.e.\ due to the influence of the Earth's magnetic field in deflecting the $e^{\pm}$ in the cascade \cite{1967Natur.215..267A}. In all other applications, the influence of the magnetic field is negligible, either due to the short paths of particles in the medium (in-ice experiments), and/or the lack of an appreciable magnetic field (the Moon).

Macroscopically, the effect of the magnetic field is to generate a transverse current, which rises and falls with the shower development. Also known as the moving dipole model, this is essentially the explanation put forth by Ref.\ \cite{1966RSPSA.289..206K}, and developed into the modern era through \citet{2008APh....29...94S} and successors. Microscopically, however, the effect appears to be the deflection of particles in the Earth's magnetic field --- i.e.\ synchrotron radiation.

Synchrotron radiation arises from the helical motion of a charged particle in a uniform magnetic field. The classical derivation of the synchrotron radiation spectrum, first given by Ref.\ \cite{PhysRev.75.1912}, is available in many textbooks (e.g.\ Refs.\ \cite{1998clel.book.....J,2007psr..book.....H}), and makes the assumption of an ultra-relativistic particle ($\beta \approx 1$). Applied to particles in an extensive air shower, which move through the Earth's magnetic field, it is known as geosynchrotron radiation \cite{2003A&A...412...19H}. 

The current consensus is that geomagnetic radiation most closely resembles the transverse current model, since the interaction length of particles in the Earth's atmosphere is less than the distance over which a pulse is observed.

\begin{figure}
    \centering
    \includegraphics{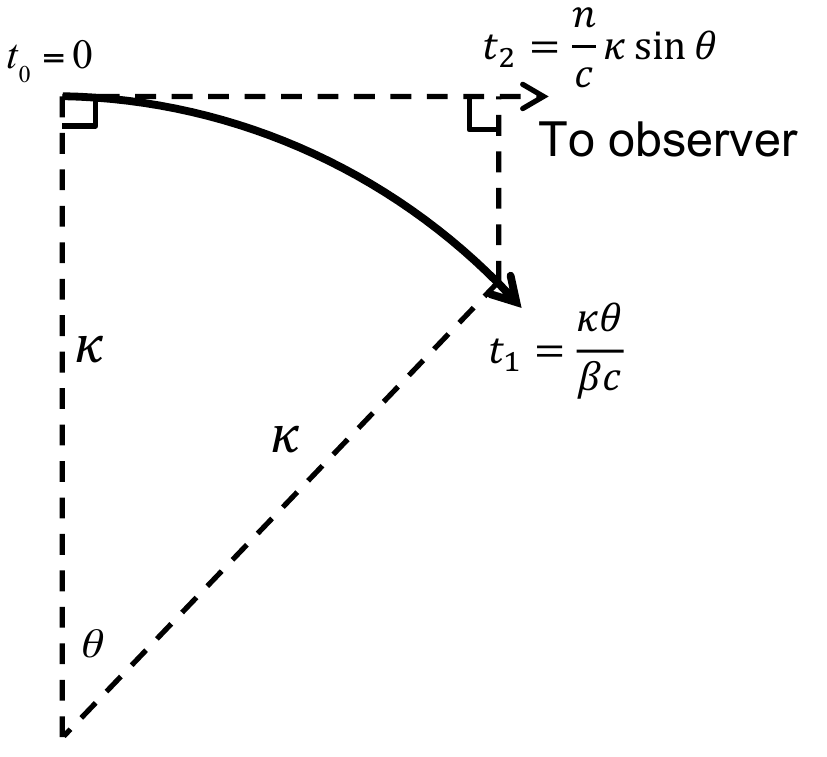}
    \caption{Sketch of the time delays associated with synchrotron radiation. At time $t_0=0$ a particle is oriented towards an observer. After subtending an angle $\theta$ with curvature radius $\kappa$ at velocity $v=\beta c$, it is time $t_1=\kappa \theta (\beta c)^{-1}$. Radiation emitted at time $t_0$ takes a time $t_2=(n/c) \kappa \sin \theta$ to be equidistant to the observer.}
    \label{fig:synch_sketch}
\end{figure}

To understand this, consider the case of an observer viewing a particle undergoing a circular arc in a medium, shown in Fig.~\ref{fig:synch_sketch}. Expanding $\sin \theta$ to $\mathcal{O}\sim\theta^3$, the time delay in emission over the arc, $\Delta t \equiv t_1-t_2$, is
\begin{eqnarray}
\Delta t = \frac{n \kappa}{c} \left[\frac{\theta^3}{6}-\theta \left(1-\frac{1}{n \beta} \right) \right]. \label{eq:synch_dt}
\end{eqnarray}
From Eq.~(\ref{eq:synch_dt}), the time delay arises from two sources: the relative velocities of the particle and the speed of light in the medium ($\theta$ term), and the curvature of the arc ($\theta^3$ term). The first is identical to that for the Askaryan effect, while the second is unique to the geomagnetic effect.

The instantaneous radius of curvature $\kappa$ is
\begin{eqnarray}
\kappa & = & \frac{r_g}{\sin^2 \alpha} \label{eq:kappa} \\
r_g & = & \frac{E \beta \sin \alpha}{|q| B c}
\end{eqnarray}
for magnetic field strength $B$, angle between the field and velocity vector $\alpha$, relativistic gyroradius $r_g$, and particle energy $E$ and charge $q$.

For coherency, I characteristically require $\Delta t < \pm (4 \nu)^{-1}$, such that the phase delay between emission at $\theta$ and $-\theta$ is less than half a wavelength. Setting $\Delta t = (4 \nu)^{-1}$ defines the characteristic angle $\theta_{\rm synch}$ over which emission is coherent, corresponding to a distance $\ell_{\rm synch} = 2 \kappa \theta_{\rm synch}$. $\ell_{\rm synch}$ sets the scale over which particle interactions would disrupt the synchrotron-like emission for an observer in the forward direction.

In the case of $n \beta =1$ (the traditional case of an ultra-relativistic particle in vaccuum), Eq.~(\ref{eq:synch_dt}) can be solved simply in terms of $\theta$. The coherency condition thus gives
\begin{eqnarray}
\theta_{\rm synch} & = & \left( \frac{3 c}{2 \nu \kappa} \right)^{\frac{1}{3}} \label{eq:theta_synch} \\
\ell_{\rm synch} & = & 2 \kappa^{2/3} \left( \frac{3 c}{2 \nu} \right)^{\frac{1}{3}} \nonumber \\
& = & \left(  \frac{2 E \beta }{|q| B \sin \alpha} \right)^{\frac{2}{3}} \left(\frac{3}{c \nu}  \right)^{\frac{1}{3}}.
\end{eqnarray}
For the atmosphere, where $1-(n \beta)^{-1} < 0.003$, the full solution is almost identical. This is not generally the case, and for $n$ appreciably greater than unity, the situation becomes more complex \cite{2015MNRAS.449..794R}.

Note that the classical result of
radiation from a relativistic source with Lorentz factor $\gamma$ being beamed into an angle $1/\gamma$ is different from a coherency consideration. The former considers how radiation emitted in the rest frame appears from the observer frame, irrespective of whether or not it maintains spatial coherency. Here, only coherency is considered, irrespective of the angular distribution of radiation. Setting $\theta_{\rm synch} = 1/\gamma$ and solving for $\nu$ leads to the derivation of the critical frequency of synchrotron radiation,
\begin{eqnarray}
\nu_{\rm crit} & = & \frac{3 c}{2 \kappa} \gamma^3,
\end{eqnarray}
above which coherency considerations cut off the emission instead of the beaming effect.

\begin{figure}
    \centering
    \includegraphics[width=0.5\textwidth]{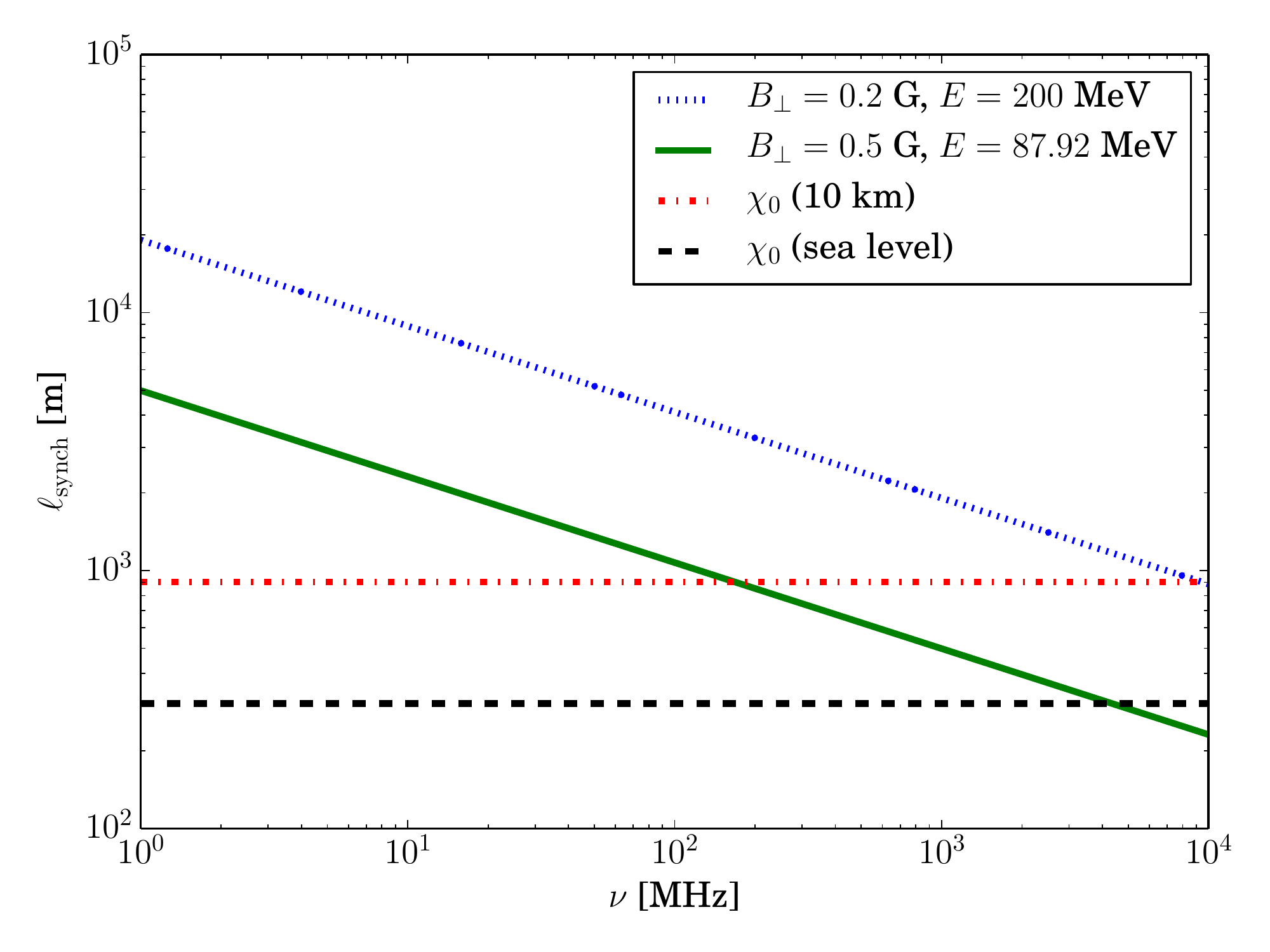}
    \caption{Characteristic length scale $\ell_{\rm synch}$ over which $e^{\pm}$ synchrotron radiation is observed for an observer in the plane of the emission. This is computed for two combinations of perpendicular magnetic field $B_{\perp}$ and electron energy $E$, and compared to the radiation length $\chi_0$ at two different altitudes.}
    \label{fig:ell_synch}
\end{figure}

Evaluating $\ell_{\rm synch}$ requires some choice for $E$ and $B_{\perp}=B \sin \alpha$. The energy $E$ of most relevance for geomagnetic radiation is $E_{\rm crit}=89$\,MeV, at which the maximum number of $e^{\pm}$ should be observed. Since an appreciable contribution will be made by particles of higher energy, a value of $E=200$\,MeV is also considered. Typical values for $B_{\perp}$ would range for $50\,\mu$T for a cascade perpendicular to the local field at most locations of radio-detection experiments, to $20\,\mu$T for a cascade at large angles to the local field, or at the Pierre Auger Observatory in Argentina, where the geomagnetic field is weaker. Fig.~\ref{fig:ell_synch} plots the expected range of $\ell_{\rm synch}$ for these values, compared to the radiation length at two different altitudes.

In regions where $\ell_{\rm synch} < \chi_0$, the emission should resemble canonical synchrotron radiation, and be accurately predicted by the geosynchrotron model. From Fig.~\ref{fig:ell_synch} however, at ground level, $\ell_{\rm synch} < \chi_0$ only for frequencies approaching $10$\,GHz for $B_{\perp}=0.5$\,G, $E=87.92$\,MeV i.e.\ only for the last generation of $e^\pm$. At $10$\,km in altitude --- a reasonable value of $X_{\rm max}$ --- the decreasing density increases the radiation length, and would allow low-energy $e^\pm$ to emit synchrotron-like radiation above $100$\,MHz. Where $\ell_{\rm synch} > \chi_0$, an overall drift in the direction of the Lorentz force will still be observed, consistent with the transverse current model.

\begin{figure}
    \centering
    \includegraphics[width=0.5\textwidth]{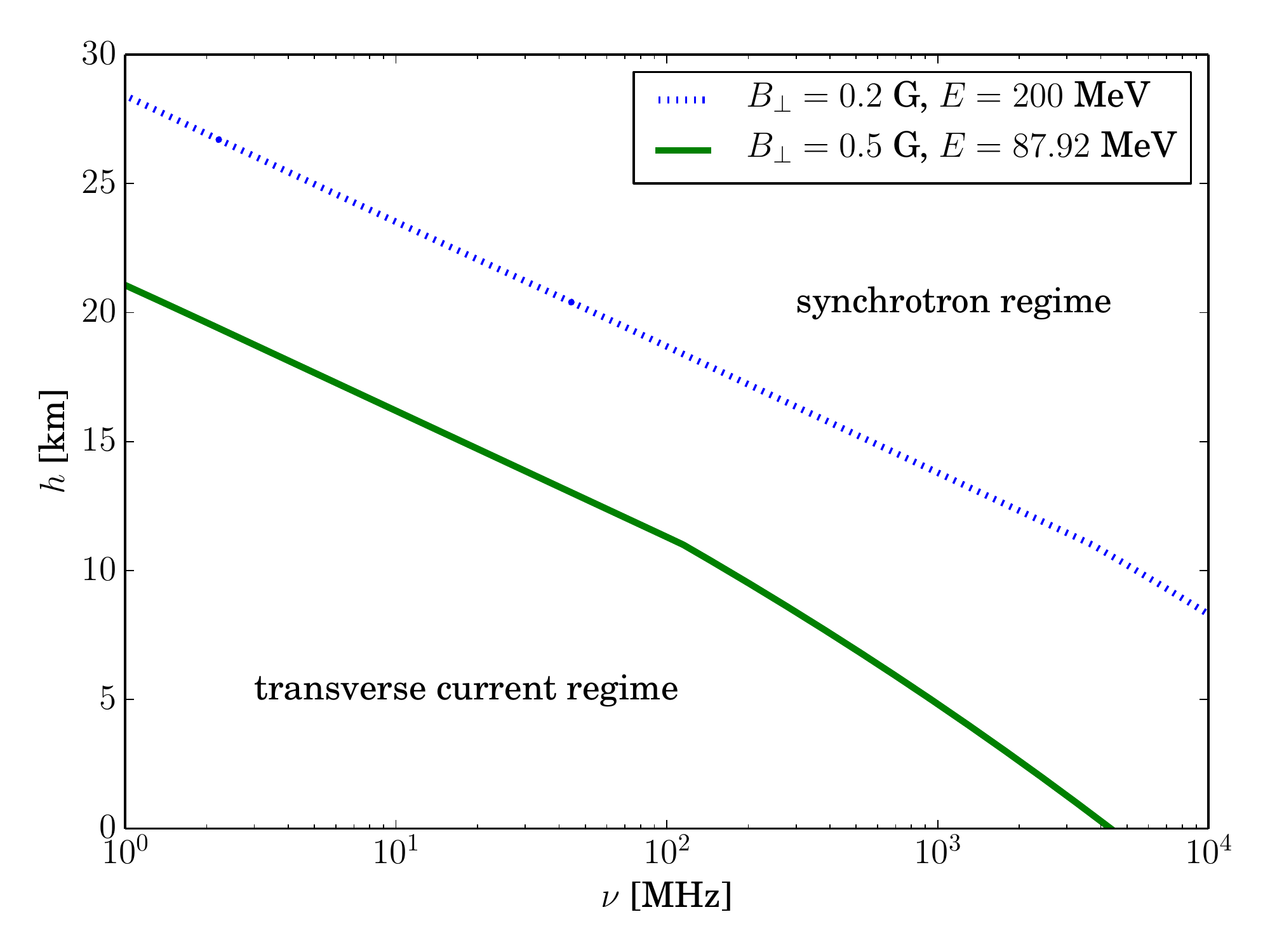}
    \caption{Height $h$ in the atmosphere at which $\ell_{\rm synch} = \chi_0$, computed for two combinations of magnetic field / electron energy. Above this height is the synchrotron regime where $\ell_{\rm synch} < \chi_0$, and below this the transverse-currentregime, where $\ell_{\rm synch} > \chi_0$.}
    \label{fig:nu_eq}
\end{figure}

Since $\chi_0$ increases monotonically with altitude, for any observation frequency, there will exist some altitude $h$ at which $\ell_{\rm synch}=\chi_0$. Above this altitude, radiation will be more synchrotron-like, and below this more transverse-current like. Again using atmospheric properties from Tab.~\ref{tab:media}, $h$ is plotted as a function of frequency in Fig.~\ref{fig:nu_eq}.

At observation frequencies below $100$\,MHz, radiation from the majority of $e^\pm$ below $20$\,km in altitude will resemble transverse current radiation. This height is typically above the majority of cascade development in EAS. This clearly explains why observations of radio-emission from EAS using ground arrays, which in the modern era have all observed below $100$\,MHz, detect radiation consistent with the transverse current model.

Simulations using \textsc{CoREAS}, when run from 3.4--4.2\,GHz for vertical cascades, do predict a more-synchrotron-like radiation pattern \cite{CoREAS}. This includes a deficit of emission in the $\vec{v} \times \vec{v} \times \vec{B}$ direction (the magnetic North-South line for vertical cascades), and a `cloverleaf' pattern in the north-south component of the electric field vector \cite{HuegeUlrichEngel2007}. While simulations using \textsc{ZHAireS} in the 300\,MHz--3\,GHz range have also been performed \cite{2012PhRvD..86l3007A}, showing that the radiated spectrum extends smoothly up to at least 1.4\,GHz at the Cherenkov angle, it is unclear if the calculations make similar predictions.

Several experiments have probed the radio emission from EAS above the $\lesssim$100\,MHz range. LOFAR observations from 110-190\,MHz have reported the expected Cherenkov ring in the ground intensity pattern \cite{2015APh....65...11N}.
ARIANNA, sensitive from 100\,MHz to 1\,GHz, reports 38 cosmic ray events with refracted signals highly correlated with \textsc{CoREAS} predictions \cite{ARIANNA_CR}. ANITA, observing in the 200\,MHz--1\,GHz range, has analysed 14 detected cosmic-ray events, finding measured signal amplitudes, arrival directions, and spectral slopes to be consistent with predictions from \textsc{ZHAireS} \cite{2016APh....77...32S}. The CROME experiment, operating from 3.4--4.2\,GHz, has detected EAS with core positions qualitatively consistent with the radiation pattern predicted by \textsc{CoREAS} \cite{CROME_2014}. None of these measurements probe the structure of events in sufficient detail to test for the changing nature of the signal however.

It may be that the most likely method to detect the synchrotron regime in EAS would be observations in the few-hundred MHz range, targeting highly inclined EAS, which develop higher in the atmosphere. While observations at very high frequencies or weak perpendicular magnetic fields will yield emission which is more synchrotron-dominated, the total signal in these cases will be intrinsically weak.

\section{Discussion}

\subsection{Scaling to full particle cascades}

The results presented so far have been obtained for very simple sources, being single particles traveling for characteristic distances in straight lines or uniform circular motion. Particle cascades are much more complicated however, with the stochastic nature of interactions, and the large spread of energies, acting to smear radio emission about expectation values characterized by radiation lengths. Furthermore, the coherent nature of radiation over the entire particle cascade means that individual particle behavior can be of secondary importance to overall cascade development. This is especially the case for Askaryan radiation, where the relevant particle motion/acceleration is in the longitudinal direction, and successive generations of particles can act as a single effective particle track. In such a case, the macroscopic analogue of bremsstrahlung is the derivative of the excess charge magnitude in the longitudinal direction, and the Askaryan component will peak where the derivative is maximum. Geomagnetic radiation however will remain proportional to the total number of charges, and peak at shower maximum. It is the offset between the effective origins --- and hence arrival times --- of these signals that leads to the circular polarisation predicted by Ref.~\cite{CoREAS} and observed by Ref.~\cite{LOFAR_polarisation}.

Ref.\ \cite{Andriga2011} model the longitudinal development of EAS using a Gaiser-Hillas profile \cite{GaisserHillas}, fitting the energy deposition $dE/dX$ as a function of atmospheric depth $X$, relative to the point of maximum development $X_{\rm max}$:
\begin{eqnarray}
\frac{dE}{dX} & = & \left(1 + R \frac{X^\prime}{L}\right)^{R^{-2}} \exp \left( - \frac{X^\prime}{R L} \right) \\
X^\prime & = & X-X_{\rm max}. \nonumber
\end{eqnarray}
Ref.\ \cite{AugerShape2019} find best-fit parameters of $0.2 \le R \le 0.27$, and $222 \le L \le 234$\,g\,cm$^{-2}$ in the $10^{18--18.2}$\,eV range (ranges are dominated by statistical uncertainties). Similar profiles would be expected for cascades in dense media. For these values, the full width at half maximum (FWHM) of the energy deposition is $540$\,g\,cm$^{-2}$, more than ten times the electromagnetic radiation length $\chi_0$.

In air (ice), this translates to a distance of 4.5\,km ($5.9$\,m), where from Fig.~\ref{fig:air_ice}, the total radiated power has a significant (dominant) contribution from particle motion, i.e.\ traditional Cherenkov radiation. Thus for a real cascade, behavior similar to both Cherenkov radiation and bremsstrahlung is expected.

In the case of geomagnetic radiation, the regime in which synchrotron radiation is expected to dominate is dependent on atmospheric height and particle energy, both of which vary greatly within a cascade. However, since the direction of particle acceleration ($q \mathbf{v} \times \mathbf{B}$) is perpendicular to the longitudinal extent of the cascade, the summed contribution of different particles within the longitudinal development cannot add to act as a single effective particle undergoing helical motion. Therefore the results of Fig.~\ref{fig:nu_eq} are expected to hold. 
Indeed, hints at the changing nature of radiation from extensive air showers at $>$\,GHz frequencies have already been seen in \textsc{CoREAS} simulations \cite{CoREAS}.

\subsection{Experimental prospects}

In this work, I have predicted that changing observation frequency will result in little change in the nature of radiation from particle cascades in dense media such as ice. However, in the case of EAS, increasing frequency from the 100\,MHz to the GHz range should produce a shift in Askaryan radiation from being bremsstrahlung-like to Cherenkov-like, and geomagnetic radiation from being transverse-current-like to synchrotron-like. How best to search for these effects?

What is required is a high-frequency (GHz) measurement of the ground pattern of an EAS. A key experimental indicator for the onset of the synchrotron regime would be the emergence of a ``clover leaf'' pattern in the North-South polarization \cite{HuegeGeosynch2007,CoREAS}, while measuring the change from almost constant to linear scaling of total radiated energy shown in Fig.~\ref{fig:air_ice}, e.g.\ using the technique of Ref.\ \cite{AugerRadioEnergyPRL}, with longitudinal extent would indicate the transition from bremsstrahlung-like to Cherenkov-like behavior. Targeting highly inclined EAS may reveal such signatures, since these events develop higher in the atmosphere, and have a longer effective tracklength over which synchrotron-like and Cherenkov-like behavior can develop.

Experiments that have observed EAS in the GHz regime however are not distributed ground arrays, and only observe each cascade from a single point in the radiation pattern \cite{ANITA_results_3rd,ARIANNA_CR,CROME_2014,EASIER_2013}. However, the radio extension to IceTop \cite{IcetopRadio}, and cosmic ray investigations using the Murchison Widefield Array \cite{MWA_phaseII} and the Square Kilometre Array \cite{SKA_EAS_ICRC_2017}, all aim to observe up to $\sim300$\,MHz. In particular, the location of IceTop near the South Magnetic Pole means that all highly inclined EAS will be traveling nearly perpendicular to the local field lines. Precise observations with these instruments may be able to detect the onset of such signatures.

Tests for reduced emission power from the Askaryan effect in dense media at low frequencies however (c.f.\ Fig.~\ref{fig:air_ice}) will be difficult. The limited space available for beam targets in laboratory experiments leads to reflections contaminating the signal \cite{SLAC_magnetic,SLACT510_detailed}, and measurements down to $100$\,MHz (where such effects would be noticeable in ice) would require a target surrounded by approximately 3\,m of homogeneous material. Using a target material with high density, but relatively low refractive index, may overcome this limitation, by allowing this effect to appear at higher frequencies, and hence be studied with a practically sized beam target.

\section{Conclusion}
\label{sec:conclusion}

The complexity of modeling radiation from particle cascades is the motivation behind the development of codes such as \textsc{ZHS}, \textsc{CoREAS} and \textsc{ZHAireS}. Using these, or advanced semi-analytic methods such as \textsc{EVA} or \textsc{MGMR3D}, is required for accurate quantitative estimates of radiation properties from particle cascades. The motivation for this work however was to qualitatively explain the underlying reasons behind the nature of the radiation predicted from these codes, and observed in experiments. This has been achieved.

I have demonstrated how and why the nature of radiation from single particle tracks changes as a function of the medium and observation frequency. The nature of Askaryan radiation is governed by the distinguishability of radiation arising from the start and end points of the track, and from the particle motion. This critically depends on the tracklength in units of the wavelength in the medium, $\ell/\lambda_n$. At typical values for air, the radiation is distinguishable over most angles, and the emission appear bremsstrahlung-like. For ice, the higher refractive index means that when the emission is distinguishable, emission from the motion of the track is also distinguishable, and the emission is more Cherenkov-like.

For curved tracks, emission will be synchrotron-like if the track curves over the characteristic duration of the synchrotron pulse. Interactions on the scale of a radiation length will prevent this for the bulk of $e^\pm$ in particle cascades for sea-level air at frequencies below 10\,GHz.
This gives a clear explanation of why Askaryan emission is more bremsstrahlung-like in air and more Cherenkov-like in ice, and why geomagnetic emission from EAS more closely resembles a transverse current than synchrotron radiation. Importantly, I have shown how this situation is expected to change as a function of observation frequency: in EAS, Askaryan (geomagnetic) emission is expected to be more Cherenkov-like (Synchrotron-like) in the GHz range than the MHz range, while in dense media, the Askaryan effect may deviate from Cherenkov-like behavior below 100\,MHz. I have proposed some experiments which could test the former effect by observing highly inclined EAS in the GHz range, and the latter using accelerator experiments.

\begin{acknowledgments}
C.W.J. thanks Frank Schr\"{o}der and Tim Huege for feedback on the article. This research made use of \textsc{gnuplot} \cite{gnuplot}, and Python libraries \textsc{Matplotlib} \citep{Matplotlib2007}, \textsc{NumPy} \citep{Numpy2011}, and \textsc{SciPy} \citep{SciPy2019}. This research was supported by the Australian Government through the Australian Research Council's Discovery Projects funding scheme (project DP200102643).
\end{acknowledgments}

\appendix

\bibliography{bibliography}

\end{document}